\documentstyle[12pt]{article}
\begin{document}

\begin{center}

{\Large\bf $ w_{1+\infty} $--type   constraints   in   two--matrix   and
Kontsevich model--different approach}

\end{center}

\begin{center}
N.L.Khviengia\\em Department  of  Physics,  Tbilisi  State
University\\,University st.,2, 380043, Tbilisi, Georgia
\end{center}

\begin{center}

ABSTRACT

\end{center}

The technique of $Q$-polinomials are used to derive the $w$-
constraints in the two-matrix and Kontsevich-like model at finite
$N$. These constraints are closed and form Lie algebra. They are
 associated with the matrices, $\lambda
^n{\partial}_\lambda^m$ with $n,m\geq 0$. In the case of
two-matrix model they can be reduced to the $W$-constraints
of \cite{8}. For the case of Kontsevich-like model and two-matrix model
with the finite polinomial potential, the number of constraints
are limited by the power of the finite matrix potential i.e. the spin
of $w$-s coincide with that power. This statement is the natural
consequence of the form of constraints.

\newpage

Recently the great deal of progress  has  been  made  in  the
nonperturbative  formulation  of  low-dimensional  toy  models  of
string theory--two-dimensional gravity coupled to $c<1$ matter--by
formulating the theory in terms of large-$N$ matrix integrals \cite{1,2}.
Soluble matrix integrals give  the  full  partition  function  for
string theory in simple backgrounds.

     One of the prominent features of the two dimensional  quantum
gravity and string in less than one dimension is the appearance of
the Virasoro or the $W$--algebra  constraints  for  the  partition
function \cite{3,4}. The time-dependent partition function is identified
with the corresponding $\tau$--function.  The  role  of  times  is
played by the coupling constants in the  matrix  potential.  These
constraints imposed on the $\tau$--function select a sub-class  of
$\tau$-functions from the entire space of generic $\tau$-functions
and may be considered as equations of motion in the general string
theory associated with the model.

     In this letter we are deriving the $w_{1+\infty}$-constraints
in Kontsevich-like(KL) and two-matrix model. Instead of orthogonal
polinomial method we  follow the  technique  formulated  in  \cite{5}.
The crucial point of this approach is that these type of constraints
is generated by ${{{\lambda}^{n}{\partial}_{\lambda}^{m}}}$ with
$n.m\geq{0}$ \cite{6}.

     The partition function of Kontsevich  model  is  proportional
to the $N\times N$ (anti)Hermitian matrix integral
\begin{equation}
{\cal F}[{\Lambda}]=\int DX \exp{-tr W[X]-tr\Lambda X}
\label{IN1}
\end{equation}
with $W[X]=X^{3}$ and $N\rightarrow\infty$.

     The importance of Kontsevich-model appearing was the sprang
possibility to describe the continuum limit of one-matrix models.

      Using  the technique  of  \cite{5}  we  obtained  the
$w$--constraints in the most  non-trivial  way.  It  turns   out,
that in the case of KL-models  i.e.  with  the  matrix  potential
of finite dimension the $w$-constraints are depending merely on the
power of potential and the spin of $w$ i.e.the number of constraints
are  limited by that power automatically.  The two-matrix- case is very much
akin of the KL-models when the potential of it is consist  of  two
pieces: polinomial of finite degree and of infinite one.

\section{One-Matrix model}
     Let  us  demonstrate  the  benefits  of   the   method   (the
detailed description of that can be found in \cite{5})   for  the
1-matrix model case. The partition function of it is following
\begin{equation}
Z_{N}({g_{l}})=\int d^{N^{2}}M \exp(\sum_{l=0}^{\infty} g_{l}
Tr M^{l})
\label{IN1M}
\end{equation}

     Rewriting this integral in terms of $M$-matrix eigenvalues,we
obtain
\begin{equation}
Z_{N}(g_l)=  \int  d^N\lambda \Delta^2 (\lambda)
\exp U(g_l)
\label{INEI1}
\end{equation}
where
\begin{equation}
U(g_l)= \sum_{l=0}^\infty \sum_{i=1}^{N} g_l
\lambda_i^l
\label{UU}
\end{equation}
Here the integration over the angular variables
have been performed and $\Delta (\lambda)= \prod _{i<j}
({\lambda}_{i}-\lambda_{j})$ --- is the Wandermonde determinant.

     The $w_{1+\infty}$--algebra due to \cite{6} can  be  generated  by
the operator
\begin{equation}
O_{nm}=x_{n} \frac{{\partial}^{m}}{\partial x^{m}}
\label{Onm}
\end{equation}
where $n,m\in {\bf Z}_+$. It is represented by the commutator
\begin{equation}
[O_n^m, O_l^k]=\{(k+1)n-(m+1)l\}O_{n+l}^{m+k}
\label{OnmCOM}
\end{equation}

     The algebra, that is isomorphic to the $w_{1+\infty}$ can  be
written in terms of eigenvalues $\lambda_i$ \cite{5}
\begin{equation}
D_{n,m}\equiv \sum_{i=1}^{n} {\lambda}_{i}^{n}
\frac{\partial^m}{\partial \lambda_i^m}
\label{Dnm1}
\end{equation}
and conjugate one
\begin{equation}
  D_{n,m}^{+}=(-1)^{m}\sum_{i=1}^{N} \frac{\partial^{m}}
  {\partial{\lambda}_{i}^{m}}{\lambda}_{i}^{n}
\label{Dnm2}
\end{equation}

     Let us define the operator
\begin{equation}
 D_{m}(p)\equiv\sum_{n=0}^{\infty}D_{n,m} \frac{1}{p^{n+1}}=
  \sum_{i=1}^{N} \sum_{n=0}^{\infty}
\frac{\lambda_i^n}{P^{n+1}}
\frac{\partial^m}{\partial\lambda_i^m}=
\sum_{i=1}^{N} \frac{1}{P-{\lambda}_{i}}
\frac{\partial^m}{\partial \lambda_i^m}
\label{DEFD}
\end{equation}

     It is ,in some sense, the generating function for $D_{n,m}$.
The following formulae are useful to repeat
\begin{equation}
 \Delta^{-1} D_s(P)\Delta=\frac{1}{s+1}(\frac{\partial}
{\partial P}+ W(P))^{s+1}{\bf 1},\quad s<N,
\label{DELTA}
\end{equation}
where
\begin{equation}
 W(P)=\sum_{i=1}^{N} \frac{1}{P-\lambda _{i}}
\label{W}
\end{equation}
by definition.

     For the further purposes we need a slightly different form of
$W(P)$. Let us mention here,  that  the  representation (\ref{W})is
singular at  $\lambda _{i}=P$. We need to reexpress the $W(P)$  so
as to be {\em regular} at that point. It can be achieved by
writing (\ref{W}) in terms of "time"  $\partial/\partial g_s$-
differentiation. We find
\begin{equation}
 W(P)=\sum_{i=1}^{N} \frac{1}{P-{\lambda}_{i}}=\sum_{s=0}^{\infty}
P^{-s-1}\frac{\partial}{\partial g_{s}}\equiv
\frac{d}{dj}(P)
\label{WRED}
\end{equation}

     The   $w$-constraints are  obtained   from   the   "partial
differentiation"  formula
\begin{equation}
 0=\int d^{N}{\lambda}_{i}{{\cal A}D_{s}(P){\cal B}-(D^+_s(P){\cal A})
{\cal B}}
\label{MAIN}
\end{equation}
where ${\cal A}=\Delta(\lambda)$ and ${\cal B}=\Delta(\lambda)
\exp(U)$.

     We can write the  following  expression  for  the  high  order
differential operator
\begin{equation}
 D_{s}(P)(\Delta \exp(U)):=\sum_{m=0}^{s}(D_{m}(P)\Delta)
(D_{s-m}(P) \exp(U))C_{s}^{m}
\label{DELEX}
\end{equation}
where $C_{s}^{m}$--are binomial coefficients.

     Let us note here,  that  for  the  finite  flat  entries  the
formula (\ref{DELEX}) is just the trivial Newton--Leibnitz  formula.The
left hand side is singular at $\lambda_i=P$ and we define it by
means of {\bf residue}. On the right  hand  side,  the  first  term
turned out to be regular  at  that  point  in  the  representation
(\ref{WRED})and the second term is singular. Hence, on the  right  hand
side the sign {\bf res} will be written(supposed)  to  the  second
term. It is found to be
\begin{equation}
D_s(P)\exp(U)\equiv res D_s(P) \exp(U)\mid_{\lambda_{i}=P}=
res\sum_{i=1}^{N} \frac{1}{P-\lambda}
\frac{\partial^s}{\partial\lambda_i^s}
\exp(U)\equiv\frac{1}{s+1}
Q_s[j(P)] \exp(U)
\label{REDEF}
\end{equation}
 where
\begin{equation}
j(p)\equiv \sum_{l=0}^{\infty} lg_{l}P^{l-1}
\label{jDEF}
\end{equation}
and $Q_{l}$ is defined as
\begin{equation}
 Q_s[f]\equiv (\frac{\partial}{\partial P}+f(P))^{s}
\label{QDEF}
\end{equation}

     Thus, after all these artificial procedures formula
(\ref{DELEX}) is transformed to the expression
\begin{equation}
 D_{s}(P)(\Delta\exp(U))=\frac{\Delta}{m+1}:Q_{s+1}[j(P)+\frac
{d}{dj}(P)]:_-\exp(U)
\label{MD}
\end{equation}

Here the sign $: :$ --means the normal ordering i.e.$\frac{d}{dj}$
standing before $j$ and "${}_-$" script  is  the  projection  to  the
negative powers of $P$.(The meaning of the latter will be  cleared
out below in considering the reduction of the constraints).

     After all the preliminary procedures we  can  return  to  the
(\ref{MAIN})and inserting formulae (\ref{DELTA}),
(\ref{MD}) into the (\ref{MAIN}) arrive to  the  following  expression
for the $w_{1+\infty}$--constraints
\begin{equation}
 w_{1+\infty}^{(l)}= \frac{1}{l}:Q_{l}[j(P)+\frac{d}{dj}(P)]:
+(-1)^{s}Q_{l}^{+}[\frac{d}{dj}(P)],\quad l<N,
\label{w1}
\end{equation}

according to the [5] $w^{(l+1)}$ are stated to be reducing to  the
Virasoo type ones. For example, for $l=2$
\begin{equation}
 w^{(2)}=(\frac{d}{dj})^{2}+(j \frac{d}{dj})_-=\sum_{l=1}^{\infty}
{\cal L}_nP^{(-n-1)}
\label{VIR}
\end{equation}
where as it is easy to guess, the ${\cal L}_{n}$ give  the  Virasoro
algebra\cite{5}. For $l=3$ the expression is the following
\begin{equation}
 w^{(3)}=(jw^{(2)})_-+\partial_{P}w^{(2)}
\label{w3}
\end{equation}
and so on.

 The important item to stress, is that the set of constraints
(\ref{w1})is closed. They form the Lie algebra; The commutator of
$lw^{(l)}$ with its conjugate is the following\footnote{In (\ref{lw})
from the left and the right hand side we took the {\em residue}
for the following purpose: In order to obtain the Lie algebra
for $lw$-s we need one and the same argument $P$ for $lw$. In that
case we should take commuting $j(P)$ and $\overline j(q)$, which
is singular when $P=q$.}
\begin{equation}
 [lw^{(l)}(P), mw^{+(m)}(P)]_={(-1)^{l}(Q_{l+m}[f]-Q_{l+m}[g])+
                  (-1)^{m}(Q_{l+m}[-f]-Q_{l+m}[-g])}
\label{lw}
\end{equation}
Taking into account (\ref{VIR}),for $l=m=2$ we obtained the Virasoro
algebra\footnote{The  commutator  of  $lw^{(l)}$  with  itself  is
giving the same result for $l$ and $m$ --even.}

\section{Two--Matrix model case}

     Let us turn to the main result of the  paper--to the  two-matrix
model. The partition function for the case is the following
\begin{equation}
 Z_{V,W}= \int DX D\Lambda \exp\{-trV[\Lambda]-trW[X]-tr\Lambda X\}
\label{PAR1}
\end {equation}

     In terms of eigenvalues the integral will have the form
\begin{equation}
 Z_{V,W}=\int \prod_{\alpha=1}^{2}d^{N}\lambda_{i}(\alpha)\Delta
(\lambda(1))\Delta(\lambda(2))\exp\{-\sum_{k=1}^{\infty}t_{k}\lambda
_{i}^{k}(1)-\sum_{m=1}^{K}g_{m}\lambda_{i}^{m}(2)+\sum\lambda_{i}
(1)\lambda_{i}(2)\}
\label{PAREI}
\end{equation}
Here we denote by ${\lambda}_{i}(\alpha)$ the eigenvalues of
accordingly--$X$ and $\Lambda$.

In the initial condition we set
\begin{equation}
{\cal A}=\Delta(\lambda_{i}(1))\Delta(\lambda_{i}(2))\exp\{-\sum_
{m=1}^{\infty}t_{m}\lambda(1)-\sum_{m=0}^{K}
g_{m}\lambda(1)\} \equiv
\Delta(\lambda(1))\Delta(\lambda(2))\exp(U)
\label{AB1}
\end{equation}
\begin{equation}
{\cal B}=\exp(\sum\lambda_i(1)\lambda_i(2))
\label{AB2}
\end{equation}
and the operator $D_{s}$ is defined as follows:
\begin{equation}
 D_{S}(P)\rightarrow {\cal D}_{l} \equiv \sum D_{n,m}
 (\lambda(1))\frac{1}{P^{n+1}}-\sum D_{m,n}(\lambda(2))
 \frac{1}{P^{m+1}}\equiv{\cal D}_{l}^{(1)}+{\cal D}_{l}^{+(2)}
 \label{cal}
\end{equation}

     Again,we have to derive the action of ${\cal D}_{l}^{+}$ on  the
$\Delta$ and $exp(U)$.
\begin{equation}
 {\cal D}_{l}^{+(1)}{\cal A}=\Delta(\lambda(2))D_{l}^{+}(P)
\Delta(\lambda(1))\exp(U)=\Delta(\lambda(1))\Delta(\lambda(2))
\frac{(-1)^l}{l+1} :Q_{l+1}^{+}
[j+\bar{j}]:\exp(U)
\label{A}
\end{equation}
where by $j_{1}$ and$\overline j_{1}$ are denoted the expressions
\begin{equation}
 j \equiv \sum^{\infty}lt_{l}P^{l-1}
\label{Dj},
 \end{equation}

\begin{equation}
j_{1}\equiv\sum^{K}lg_{l}P^{l-1}
\label{Dj1},
 \end{equation}

\begin{equation}
\bar{j}\equiv \frac{d}{dj_{1}}\equiv \sum^{\infty}
P^{-l-1}\frac{\partial}{\partial t_{l}}
\label{jD1},
 \end{equation}

\begin{equation}
\bar{j}_{1}=\sum^{K}P^{-l-1}\frac{\partial}{\partial g_{l}}
\label{jD2}
 \end{equation}
and
\begin{equation}
-{\cal D}_{l}^{+(2)}{\cal A}= \Delta(\lambda(2))\Delta(\lambda(1))
:\frac {Q_{l+1}^+}{l+1}[\bar{j}_1+j_1]:\exp(U)
\label{D1}
\end{equation}

Let us emphasize here, that $D_{l}^{(2)}$ depends on $l$ through
differentiation and its action on $\sum_{m=0}^{K} g_{m}P^{m}$
is zero for $l>K$. Hence, the power of $Q_{l,K}$ is limited by $K$.

     The $w$--constraints in this case is found to be
\begin{equation}
w_{1+\infty}^{(l,K)}=\frac{1}{l}:Q_{l,K}^{+}[\bar{j}_1+j_1]:_-
+:Q_l^+[j+\bar{j}]:{}_-.
\label{W2}
\end{equation}

These constraints are closed(see the commutator (\ref{lw}) and
form the Lie algebra. They can be written in terms of "times"
for spin-2 and-3. The spin-2 case constraints are the following
\begin{equation}
 w_{2}\propto (\frac{d}{dj})^{2}+(j\frac{d}{dj})=\sum\overline
{\cal L}_{n}P^{-n-1}
\label{WWW2}
\end{equation}
(see \cite{8} for the $\cal L$ definition)
and for spin-3
\begin{equation}
w_{3}\propto Q_{3,K}^{+}+Q_{3}^{+}\sim W^{(3)}
\label{WWW3}
\end{equation}
We can easily check, that \ref{WWW3} is giving the $\tilde{W}^{(3)}$
that of \cite{8}.

\section{The Kontsevich-like model}
     The KL-model is invented to define the  continuum  limit  of
the 1-matrix model. It is  still  fascinates  the  theoreticians,
since it is describing the topological gravity in the case of
$W[X]=X^{3}$. The partition function for this model  has  a  form
(\ref{IN1}). Where
\begin{equation}
W[X]=\sum_{m=0}^{K} g_{m}X^{m}
\label{FAS}
\end{equation}

     In rewriting the integral in terms  of  eigenvalues,we  need
diagonalization of $X$
\begin{equation}
X\rightarrow MX_{D}M^{+}
\label{DIA}
\end{equation}
where the $X_{D}$ is the diagonal matrix, with the eigenvalues
$\lambda_1,\ldots , \lambda_N$ and $DX$ is the Haar measure
\begin{equation}
DX=DM\prod_i^N d\lambda_i\Delta^2(\lambda)
\label{MEAS}
\end{equation}
while $W[X]$ is invariant under the transformation M
\begin{equation}
MW[X]M^{+}=W[X]
\label{M}
\end{equation}

     Using the expression for the Itsikson-Zuber integral we  are
arriving  to  the  KL-model  partition  function   in   term   of
$\lambda_{i}$
\begin{equation}
Z_{W}= \int \prod d{\lambda}_{i}\frac{\Delta({\lambda}_{i}(1))}
{\Delta({\lambda}_{i}(2))} \exp\{-\sum_{l=0}^{K}g_{l}{\lambda}_{i}
^{l}-\sum_{i}{\lambda}_{i}(1){\lambda}_{i}(2)\}
\label{IZZ}
\end{equation}

     Inserting this equation into (\ref{MAIN}) and taking
\begin{equation}
{\cal A}= \frac{\Delta(\lambda(1))}{\Delta(\lambda(2))}
\exp(W)
\label{CLA},
\end{equation}

\begin{equation}
{\cal B}= exp(\sum {\lambda}_{i}(1){\lambda}_{i}(2))
\label{CLB}
 \end{equation}

   It is important to mention here, that the operator (\ref{cal})
is not producing the "multiloopcolor"  operator  i.e.  the  cross
term like $\lambda(1)\lambda(2)$. Hence, the only remaining  term
in (\ref{MAIN}) will be $({\cal D}_{l}^{+}\cal A)\cal B$
\begin{equation}
{\cal D}_{l}^{+(1)}(P){\cal A}=\frac{(-1)^l}{(l+1)}\frac{\Delta(\lambda(1))}
{\Delta(\lambda(2))}:Q_{l+1}^{(+)}[j_{1}(P)+\bar{j}_{1}(P)]:_-\exp(U)
\label{Dl1}
\end{equation}
(For the the details of the calculation see {\bf Appendix})
and for the second part
\begin{equation}
-{\cal D}_{l}^{+(2)}{\cal A}=\frac{\Delta(\lambda(1))}
{\Delta(\lambda(2))}\{\frac{1}{2}\partial_{P}^{(s-1)}Q_{2}[\bar{j}(P)]+
\frac{1}{4}\partial_{P}^{(s-2)}Q_{2}^{2}[\bar{j}(P)]\}\exp(U)
\label{Dl2}
\end{equation}

     The $w$-constraints becoming as
\begin{equation}
w^{(l+1)}_{1+\infty}=\frac{(-1)^{l+1}}{l+1}:Q_{l+1}^{(+)}[j_{1}+\bar{j}_{1}]+
\frac{1}{4}\partial_{P}^{(s-2)}Q_{2}^{2}[\bar{j}_{1}]-\frac{1}{2}
{\partial}_{P}^{(s-1)}Q_{2}[\overline j_{1}].
\label{wMN}
\end{equation}

 As a conclusion, we stress the main points of the article;
We derive the $w$-constraints for the two-matrix model case for the
two pieces of potential: infinite and finite. In this case within
the approach, the constraints depend merely on the power of the finite
-part potential, thus restricting automatically the number of them
or equivalently, the spin of $W$-constraints. It is important to
stress here once more, that the restriction of the spin of $w$ is
straightforward, since $Q_s$ depend merely on the power of
the potential. This suggestion was first put
by E.Gava and K.Narain \cite{7} and demonstrated later by A.Marshakov,
A.Mironov and A.Morozov \cite{8,9}. The obtained constraints form the Lie
algebra and can be reduced to that of \cite{8}, when written in terms
of "time". We find the $w$-constraints for the Kontsevich-like model
with finite potential i.e. the discrete analogue of the initial
Kontsevich model.

We acknowledges S.Kharchov, Yu.Makeenko and A.Morozov for useful
discussions.

\newpage

\section*{Appendix}
 We give here the the method of calculating the action of $D_{s}(P)$
on the ${\Delta}_{-1}$. As an example, consider the case of $D_{3}(P)$

$$\Delta D_{3}(P){\Delta}^{-1}=\hbox{res}\{\sum\frac{1}{P-\lambda_j}
(-\partial_{\lambda_j}^{2}(\Delta^{-1}\partial_{\lambda_j}\Delta)+
\partial_{\lambda_j}({\Delta}^{-1}{\partial}_{{\lambda}_{j}}
\Delta)^{2})\} \eqno{(A.1)}$$

Here we have used the well-known relation
$$\Delta D_{1}(P){\Delta}^{-1}=-{\Delta}^{-1}D_{1}(P)\Delta
  \eqno{(A.2)}$$
It is important to stress here, that the action of $D$ on ${\Delta}^{-1}$
is defined through {\em res}\footnote{In this case in the (14) in the
right hand side both  terms are singular and the $\hbox{res}$ sign
there is implied for both parts by definition.} i.e. on
the right and left hand side is assumed $hbox{res}$. Using the
equation of \cite{5} we obtain

$$
{\Delta}^{-1}{\partial}_P^s\Delta = \hbox{res}(\Delta^{-1}D_s(P)\Delta)=
\frac{1}{s+1}Q_{s+1}[W(P)]
\eqno{(A.3)}$$
Thus, inserting (A3) into (A1) we obtain for the right
hand side of (A1)

$$\Delta D_3(P)\Delta^{-1}=\frac{1}{4}Q_{2}^{2}-
\frac{1}{2}{\partial}_{P}Q_{2}
\eqno{(A.4)}$$

By induction we can find the expression for ${\Delta}D_{s}{\Delta}
^{-1}$  for $\forall s$

$$\Delta D_s(P)\Delta^{-1}=\frac{1}{4}{\partial}_{P}^{s-2}Q_{2}^{2}-
\frac{1}{2}{\partial}_{P}^{s-1}Q_{2}
\eqno{(A.5)}$$

\end{document}